\documentclass[aps,prb,twocolumn,showpacs,superscriptaddress]{revtex4-1}
\usepackage[colorlinks ,linkcolor=blue,anchorcolor=blue,citecolor=blue,urlcolor=blue]{hyperref}
\usepackage{graphicx}
\usepackage{epstopdf}
\usepackage{textcomp}
\usepackage{diagbox}
\def\be{\begin{equation}} \def\ee{\end{equation}}
\def\bea{\begin{eqnarray}} \def\eea{\end{eqnarray}}

\begin{document}

\preprint{APS/123-QED}

\title{Gate-tunable chiral phonons in low-buckled group-IVA monolayers}

\author{Hanyu Wang}
\affiliation{School of Physics and Technology, Nanjing Normal University,Nanjing 210023, China}
\affiliation{Center for Quantum Transport and Thermal Energy Science(CQTES), NNU-SULI Thermal Energy Research Center (NSTER), Nanjing Normal University,Nanjing 210023, China}

\author{Zhichao Zhou}
\email{zczhou@njnu.edu.cn}
\affiliation{School of Physics and Technology, Nanjing Normal University,Nanjing 210023, China}

\author{Hao Chen}
\affiliation{School of Physics and Technology, Nanjing Normal University,Nanjing 210023, China}
\affiliation{Center for Quantum Transport and Thermal Energy Science(CQTES), NNU-SULI Thermal Energy Research Center (NSTER), Nanjing Normal University,Nanjing 210023, China}

\author{Chongqun Xia}
\affiliation{School of Physics and Technology, Nanjing Normal University,Nanjing 210023, China}
\affiliation{Center for Quantum Transport and Thermal Energy Science(CQTES), NNU-SULI Thermal Energy Research Center (NSTER), Nanjing Normal University,Nanjing 210023, China}

\author{Lifa Zhang}
\email{phyzlf@njnu.edu.cn}
\affiliation{School of Physics and Technology, Nanjing Normal University,Nanjing 210023, China}
\affiliation{Center for Quantum Transport and Thermal Energy Science(CQTES), NNU-SULI Thermal Energy Research Center (NSTER), Nanjing Normal University,Nanjing 210023, China}

\author{Xiao Li}
\email{lixiao@njnu.edu.cn}
\affiliation{School of Physics and Technology, Nanjing Normal University,Nanjing 210023, China}
\affiliation{Center for Quantum Transport and Thermal Energy Science(CQTES), NNU-SULI Thermal Energy Research Center (NSTER), Nanjing Normal University,Nanjing 210023, China}

\date{\today}

\begin{abstract}

We investigate the electric response of chiral phonons on the low-buckled group-IVA monolayers by performing first-principles calculations.
The vertical electric field breaks the degeneracy of phonon modes at high-symmetry $\pm K$ points of the phonon Brillouin zone,
 and the size of the phononic gap is proportional to the strength of the electric field.
  The gapped phonon modes at $\pm K$ possess chiralities with considerable phonon circular polarizations
  and discrete phonon pseudoangular momenta.
  The chiralities of phonons are robust against the variation of the field strength,
but reversed by changing the field direction.
Electric control of chiral phonons adds a new dimension to the study of chiral phonons,
 which has potential use in the design of phononic and valley devices.
\end{abstract}

\maketitle

\section{Introduction}

In condensed matter physics,
spatial inversion symmetry breaking usually endow chirality to Bloch electrons and other quasiparticles.
For electronic systems,
chiralities of valley electrons have been observed in transition metal dichalcogenide (TMDC) monolayers
 through optoelectronic measurements  \cite{xiao2007,yao2008,xiao2012,schaibley2016}.
 Recently, chiral phonons with atomic circular vibrations have been proposed theoretically in honeycomb lattice with broken inversion symmetry  \cite{zhang2015}
and verified experimently in TMDC by optical pump-probe techniques  \cite{zhu2018}.
They contribute extra angular momenta besides well-known spin/orbital angular momentum  \cite{zhang2014,hamada2018},
participate in intervalley scatterings of electrons  \cite{zhu2018,zhang2020,xu2018,chen2019}
or couplings with other quasiparticles  \cite{li2019a,li2019b,guo2019,grissonnanche2020,li2020},
and enable phonon-driven topological states  \cite{liu2017},
which exhibit conceptual significance and application potential.

After chiral phonons become recognized,
a consequent intriguing question is how to selectively acquire phonons with opposite chiralities.
The selection of chiralities introduces a tunable degree of freedom to chiral-phonon-involved physical processes
and also facilitates information encoding  \cite{zhu2018,zhang2020,li2019a,li2019b,guo2019,grissonnanche2020,li2020}.
Chiralities of phonons usually originate from pristine materials without inversion symmetry in their crystalline point groups  \cite{zhu2018}
or structural symmetry reduction by e.g. proximity effect  \cite{gao2018},
thus it is hardly to tune chiralities in a controllable way.
On the other hand, compared with bulk materials,
the properties of monolayer materials can be easily modulated by applying vertical electric field  \cite{sui2015,shimazaki2015}.
 Specific to low-buckled group-IVA monolayers  \cite{cahangirov2009,liu2011a,liu2011b,drummond2012,peng2016},
 there are two atomic layers that feel different static electric potentials under vertical electric fields,
 where the inversion symmetry is broken even though the structural change is absent.
 Accordingly,
 the degeneracies, band topology and spin polarizations in their electronic structures can be modulated by vertical electric field  \cite{ezawa2012,tsai2013}.
 By analogy to electronic systems,
 it is promising to use electric fields to control phonons and their chiralities,
 which is worth studying carefully.

In this paper, we use first-principles calculations to investigate the electric field effects on chiral phonons of
low-buckled group-IVA monolayers, where silicene, germanene and stanene are considered.
For degenerate branches, phononic gaps are opened and increased with external electric fields.
Gapped phonon modes with atomic circular motions are computed to have considerable phonon circular polarizations and discrete phonon pseudoangular momenta.
 Moreover, when turning around the direction of electric fields, the chiralities of phonons are also reversed.
Our work provides a new route to control chiral phonons conveniently.

The rest of this paper is organized as follows.
In Sec. \ref{sect2_com}, we provide the details on density functional theory calculations of low-buckled group-IVA monolayers.
Definitions and the characterization methods of chiral phonons are introduced as well.
In Sec. \ref{sect3_result},
we investigate the influence of electric fields on crystal structures, phonon dispersions and vibration modes,
demonstrating the tunability of phonon chiralities.
Conclusions are made in Sec. \ref{sect4_conclu}.

\section{Computational details}
\label{sect2_com}

\subsection{Density functional theory calculations}
\label{sect2a}

Density functional theory calculations of the low-buckled group-IVA monolayers are performed
using projector augmented wave method  \cite{blochl1994} as implemented in Vienna Ab initio Simulation Package  \cite{kresse1996}.
The generalized gradient approximation of Perdew-Burke-Ernzerhof functional  \cite{perdew1996}
 is adopted to describe the exchange and correlation effects of electrons.
The energy cutoff is set to 500 eV.
 To avoid interactions between the monolayer and its periodic images,
 a vacuum space larger than $24$\ \AA\ is inserted.
A Monkhorst-Pack \textit{k}-mesh of $17\times17\times1$ is used in structural relaxations.
As for the calculations of phonon properties, we employ the finite displacement method  \cite{alfe2009} using PHONOPY package  \cite{togo2008},
where a $5\times5\times1$ supercell and a $5\times5\times1$ \textit{k}-mesh are adopted.

\subsection{Characterizations of chiral phonons}
\label{sect2b}

To illustrate the chirality of phonons conveniently,
the basis set to describe phonon eigenvectors is transformed from the Cartesian coordinate to a new orthogonal coordinate.
For low-buckled group-IVA materials, there are two atomic sites labeled by A and B in a unit cell, as shown in Fig. \ref{P1}(a).
As a result, the new basis set contains six unit vectors, i.e.
$|R_{A}\rangle\equiv\frac{1}{\sqrt 2}(1, i, 0, 0, 0, 0)^{T}, |L_{A}\rangle\equiv\frac{1}{\sqrt 2}(1,-i, 0, 0, 0, 0)^{T},|Z_{A}\rangle\equiv(0, 0, 1, 0, 0, 0)^{T},
|R_{B}\rangle\equiv\frac{1}{\sqrt 2}(0, 0, 0, 1, i, 0)^{T},|L_{B}\rangle\equiv\frac{1}{\sqrt 2}(0,0,0, 1, -i, 0)^{T},|Z_{B}\rangle\equiv(0,0,0,  0, 0, 1)^{T}$,
where $|R_{A(B)}\rangle$, $|L_{A(B)}\rangle$, and $|Z_{A(B)}\rangle$ represent the unit vectors of right-handed circular motion,
left-handed circular motion and the $z$-direction linear motion for atomic vibrations on a $A(B)$ sublattice, respectively.
A phonon eigenvector $|\xi\rangle$ can be represented in this new basis set as,
\begin{eqnarray}
  |\xi\rangle=\sum_{\alpha=A,B}(C_{R_{\alpha}}|R_{\alpha}\rangle+C_{L_{\alpha}}|L_{\alpha}\rangle+C_{Z_\alpha}|Z_{\alpha}\rangle),
\end{eqnarray}
where $C_{R_{\alpha}}=\langle R_{\alpha}|\xi\rangle$,
$C_{L_{\alpha}}=\langle L_{\alpha}|\xi\rangle$,
$C_{Z_{\alpha}}=\langle Z_{\alpha}|\xi\rangle$.
Considering the normalization condition of $|\xi\rangle$,
$|C_{R_{\alpha}}|^2+|C_{L_{\alpha}}|^2+|C_{Z_{\alpha}}|^2$ should be equal to $1$.

Then we define the operator of phonon circular polarization along the $+z$ direction as,
\begin{eqnarray}
  \hat{S^{z}}\equiv  \sum_{\alpha=A,B}(|R_{\alpha}\rangle\langle R_{\alpha}|-|L_{\alpha}\rangle\langle L_{\alpha}|).
\end{eqnarray}
The chirality of a phonon eigenvector $|\xi\rangle$ can be characterized by the phonon circular polarization,
 \begin{eqnarray}
  S^{z}_{ph}=\langle\xi|\hat{S^{z}}|\xi\rangle\hbar=\sum_{\alpha=A,B}(|C_{R_{\alpha}}|^{2}- |C_{L_{\alpha}}|^{2})\hbar.
\end{eqnarray}
The module of $S^{z}_{ph}$ represents the magnitude of chirality, and its sign represents the direction of chirality.
According to the definition,  $S^{z}_{ph}$ is exactly $+\hbar$ $(-\hbar)$ for a phonon mode with fully right-handed (left-handed) chirality.

\section{Results and discussions}
\label{sect3_result}

\subsection{Crystal structures}
\label{sect3a}

\begin{figure}[htbp]
\centerline{\includegraphics[width=8.5cm]{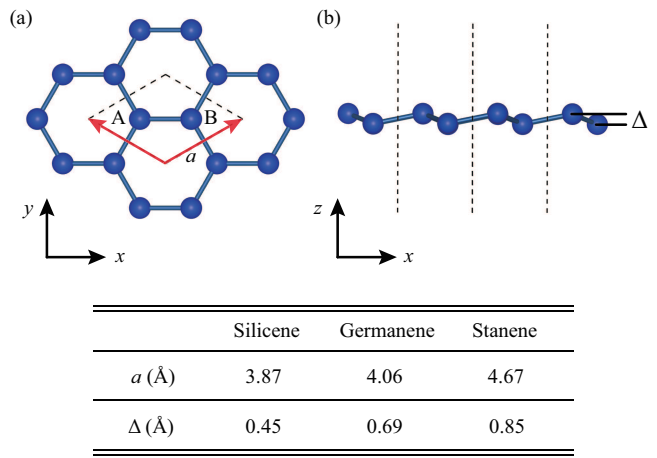}}
\caption{\label{P1} Crystal structure of a low-buckled group-IVA monolayer.
The top view and the side view are given in (a) and (b), respectively.
The dashed box represents one unit cell,
and blue balls denote group-IVA atoms.
The computed lattice constants $a$ and height differences $\Delta$ of three group-IVA monolayers after structural relaxation are also listed.
}
\end{figure}

Among the two-dimensional group-IVA materials, the crystal structures of silicene, germanene, and stanene
considered here have honeycomb lattice, as illustrated in top view of Fig. \ref{P1}(a).
 In side view of Fig. \ref{P1}(b), two atomic sites in a unit cell are located in two atomic layers of different heights along the $z$ axis,
 exhibiting a buckled structure.
 The buckled honeycomb structure has a crystalline point group of $D_{3d}$,
 in contrast to flat graphene monolayer with $D_{6h}$ group  \cite{ribeiro2015}.
 We computed lattice constant, $a$, and height difference, $\Delta$, of buckled atomic layers for three monolayers in the absence of electric field,
which are listed in Fig. \ref{P1} and agree with previous calculation results  \cite{cahangirov2009,liu2011a,liu2011b,drummond2012,peng2016}.
From silicene to stanene, the bigger is the atomic number, the larger is the lattice constant and the height difference.
It is also seen that buckling degree gradually becomes larger
with increasing ratio of $\Delta$ to $a$, i.e. $0.12$, $0.17$, and $0.18$ for three monolayers.
For graphene with $sp^2$ bonds, the ratio is zero,
while $sp^3$ hybridization leads to a ratio of $0.20$.
Therefore, the bonds of these three monolayers are regarded as a mix of $sp^2$ and $sp^3$ hybridizations  \cite{liu2011a,liu2011b}.

We also computed the fully relaxed geometric structures of three monolayers, under vertical electric fields up to $0.2$\ V/\AA\ .
 Compared with pristine monolayers, the deviations of the lattice constant and the height differences
  are less than 1\textperthousand\ and 3\%, respectively.
  Therefore, it is found that the influence of electric fields on geometric structures is tiny.

\subsection{Phononic gap opening at $K$ point under electric fields}
\label{sect3b}

\begin{figure}[htbp]
\centerline{\includegraphics[width=8.5cm]{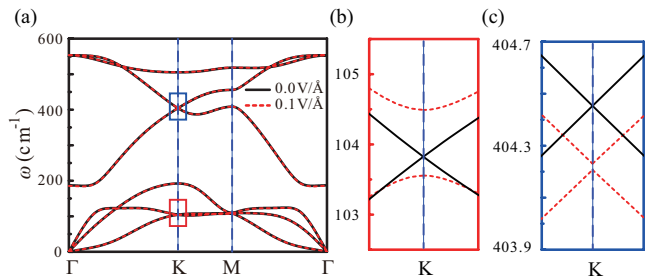}}
\caption{\label{P2} Phonon dispersions of silicene before and after applying applied electric field.
(a) The whole phonon dispersion along the path of $\Gamma-K-M-\Gamma$.
Magnified phonon dispersions around $K$ for (b) acoustic branches and (c) optical branches.
Black solid lines represent the case without applied electric field,
 while red dashed lines denote the case with an electric field of $E_z=0.1$ V/\AA\  .
}
\end{figure}

Taking silicene as a prototype of low-buckled group-IVA monolayer,
Fig. \ref{P2}(a) shows its phonon dispersion without electric field,
which is presented by black solid lines.
The phonon dispersions of germanene and stanene can be found in \ref{appendix},
which is similar to that of silicene.
There is no imaginary frequency in phonon dispersions, indicating that the crystal structure is stable thermodynamically.
At $\Gamma$ point, there are three degenerate acoustic branches with zero frequency and three optical branches above them.
For optical branches, they can be divided into two groups, i.e.
one branch with lower frequency of 185 cm$^{-1}$ and two degenerate branches with higher frequency of 553 cm$^{-1}$.
At the high-symmetry $K$ point, there are two doubly-degenerate modes, and two nondegenerate modes.
In the following part, we focus on the degenerate modes,
which belongs to acoustic branches with frequency of 104 cm$^{-1}$
and optical branches with frequency of 404 cm$^{-1}$, respectively.

Under vertical electric fields, the phonon dispersion of silicene monolayer is shown by dashed red lines in Fig. \ref{P2}(a).
The imaginary frequency is also absent in phonon dispersions, and the crystal structure is still stable.
The phonon dispersion has no obvious changes,
 except for previous degenerate points at $K$.
Figs. \ref{P2} (b) and (c) zoom in on rectangular boxes of Fig. \ref{P1}(a),
showing the evolution of phonon dispersions around $K$ for acoustic branches and optical branches that we focus on.
 Both doubly degenerate branches at $K$ point are splitted in frequency with gap openings.
   Under a vertical electric field of $E_z=0.1$ V/\AA ,
  the phononic gaps of acoustic and optical branches are 0.93 cm$^{-1}$ and 0.03 cm$^{-1}$, respectively.
  The appearance of phononic gap opening at $K$ is similar to the case of electronic band structures under applied electric field.
 It is the spatial inversion symmetry breaking under the action of applied electric field
 that lifts the degeneracy in the spectra  \cite{ezawa2012,tsai2013}.

\begin{figure}[htbp]
\centerline{\includegraphics[width=8cm]{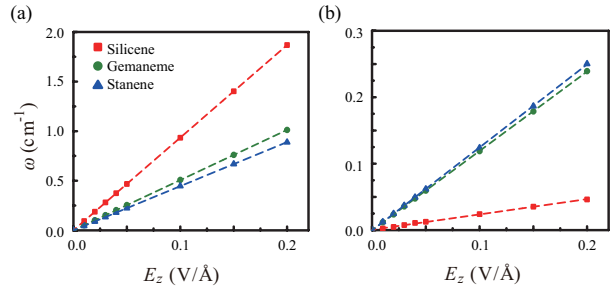}}
\caption{\label{P3}Evolutions of phononic gaps with the field strength for three group-IVA monolayers.
(a) Acoustic phononic gaps; (b) Optical phononic gaps.
}
\end{figure}

Figs. \ref{P3} (a) and (b) demonstrate the evolution of phononic gaps with the electric field strength for two branches, respectively,
where three monolayers are all considered.
These gaps are linearly proportional to applied electric field,
that is, the linear relationship is independent of monolayers and phonon branches.
Moreover, for acoustic branches in Fig. \ref{P3} (a), the ratio of the gap to the electric field strength
 decreases with increasing atomic numbers.
In contrast, for optical branches in Fig. \ref{P3} (b), the dependence between ratios and atomic numbers is reversed.
Comparing gaps between acoustic branches and optical branches,
the former is much larger than the latter one.
 Besides, different from $K$ point, the degeneracies at $\Gamma$ are unchanged.

\subsection{Gate-tunable chiral phonons at $K$ point}
\label{sect3c}

\begin{figure}[htbp]
\centerline{\includegraphics[width=8.5cm]{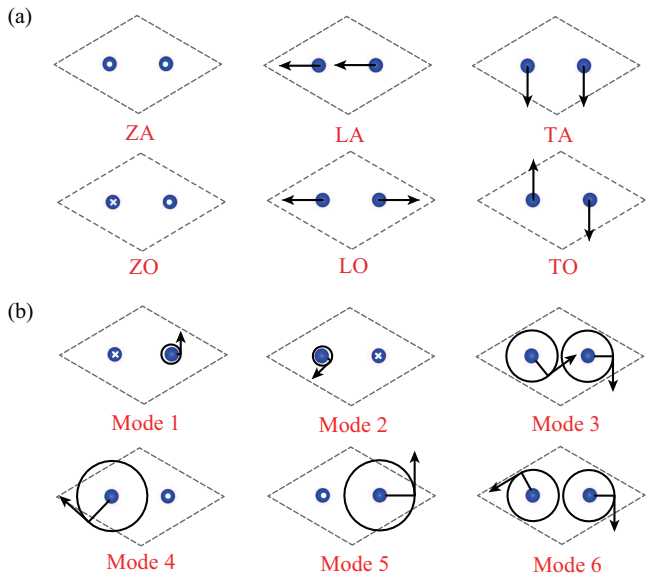}}
\caption{\label{P4}
Atomic vibrations in a unit cell for phonon modes under an electric field.
(a) Phonon modes at $\Gamma$. (b) Phonon modes at $K$.
Both (a) and (b) include six panels, corresponding to atomic vibrations of different phonon modes.
For each panel, two blue balls represent silicon atoms in a unit cell.
The radius of the circular trajectory around an atom is proportional to the amplitude of atomic circular vibration.
The arrow on the circular trajectory denotes the direction.
The point and cross on an atom represent opposite linear vibrations along the $z$ axis.
 }
\end{figure}

We then investigated atomic vibrations of different phonon modes and their chiralities under vertical electric fields.
Fig. \ref{P4} (a) shows vibrations of six phonon modes at $\Gamma$ in a unit cell of silicene.
For each phonon mode, both atoms in a unit cell are linearly vibrated.
 As for acoustic modes, interatomic relative motions are absent, while they are present for optical modes.
Three acoustic (optical) modes are further distinguished by three directions of atomic vibrations, i.e.
 out-of-plane direction, longitudinal in-plane direction along the bond, and transverse in-plane direction.
According to their atomic vibrations above,
 six phonon modes can be named as $ZA$, $LA$, $TA$, $ZO$, $LO$, $TO$.
Here,  `$Z$', `$L$', and `$T$' are abbreviations for
out-of-plane,
longitudinal,
and transverse
 vibrations, respectively.
Acoustic and optical modes are denoted by `$A$' and `$O$'.
Moreover,
the frequency of degenerate LO and TO modes is much higher than that of ZO mode,
since the in-plane bonding strength is stronger than out-of-plane one due to the two-dimensional structure of silicene.
To characterize the chirality of the phonon modes at $\Gamma$, we computed the phonon circular polarization, $S_{ph}^z$, according to its definition in Sec. \ref{sect2b}.
It is found that $S_{ph}^z$ is zero for all six modes,
indicating that these linear vibrations at $\Gamma$ have no chirality.

Moving on the $K$ point, corresponding atomic vibrations of six phonon modes are listed in Fig. \ref{P4} (b).
 It is seen that the circular atomic vibrations appear besides linear ones.
Since the above nomenclature of phonon modes at $\Gamma$ is not appropriate for the modes at $K$,
we differ these phonon modes at $K$ by their frequencies in the ascending order for simplicity.
  The third and sixth modes only possess circular vibrations in $xy$ plane.
For these two modes,
two atoms in a unit cell are clockwisely and anticlockwisely vibrated, respectively,
but with same amplitude.
 Their computed $S_{ph}^z$ are zero as well, even though circular vibrations are present.
 It is because the polatization cancellation of opposite circular motions of two atoms in a unit cell.

\begin{figure}[htbp]
\centerline{\includegraphics[width=8.5cm]{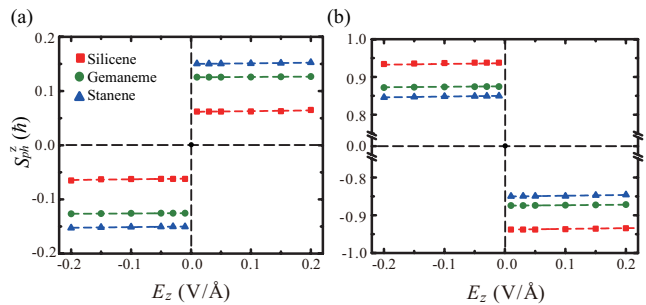}}
\caption{\label{P5} Evolutions of phonon circular polarizations with applied electric fields for three group-IVA monolayers.
(a) Lower gapped mode of acoustic branches.
(b) Lower gapped mode of optical branches.
The above two modes correspond to the first and fourth modes in Fig. \ref{P4}.
}
\end{figure}

 More interestingly,
for the rest modes at $K$ corresponding to previous gapped branches,
one atom in a unit cell exhibits linear vibration along the $z$ axis,
while the other is circularly vibrated.
For gapped acoustic branches,
 the circular motions are opposite in the first and second modes.
The same applies to the fourth and fifth modes of gapped optical branches.
Comparing with those of acoustic modes,
the amplitudes are larger for circular vibrations of optical modes.
Based on the above atomic vibrations,
the $S_{ph}^z$ under different electric fields were computed,
and their values can be found in Fig. \ref{P5}.
The computed $S_{ph}^z$ are no longer vanishing, with considerable values,
thus corresponding phonon modes exhibit definite chiralities.
Under a certain electric field,
 gapped modes have opposite circular polarizations for both acoustic and optical branches.
Therefore, for gapped modes,
only the branch with lower frequency is shown in Fig. \ref{P5}, with (a) and (b) corresponding to acoustic and optical cases, respectively.
Comparing with those of acoustic branches,
gapped modes of optical branches have much larger the magnitude of $S_{ph}^z$ due to their stronger amplitudes of circular motions.

We then concentrate on the evolutions of phonon circular polarizations with vertical electric fields for silicene monolayer,
where gapped modes of acoustic and optical branches present similar trends.
When applying electric fields along $+z$ direction,
it is seen that changes of $S_{ph}^z$ is small with an order of $10^{-3} \hbar$,
as the field strength is enhanced.
Turning around the direction of electric fields,
the magnitude of $S_{ph}^z$ is unchanged,
but the sign of it is reversed spontaneously.
This provides a effective route to select the chiralites of phonons by applied electric field.
Besides,
the gapped modes of germanene and stanene possess chiralities as well.
The evolutions of their circular polarizations, denoted by green circles and blue triangles in Fig. \ref{P5}, are also similar to the case of silicene,
 but the values of $S_{ph}^z$ are different.

Given that the three-fold rotational symmetry is still present under vertical electric field,
  pseudoangular momentum (PAM) is well-defined,
 and it was also computed for phonon modes at $K$,
 by applying three-fold rotational operator on phonon eigenvector
 and extracting the phase winding  \cite{zhang2015}.
 Table \ref{T1} gives values of PAM for six phonon modes.
 In contrast to $S_{ph}^z$, the PAM is an integer based on modulo $3$.
However, similar to $S_{ph}^z$, the PAM is insensitive to the strength of electric field,
but the signs of PAM become opposite by reversing the field direction.
 The same applies to the other two group-IVA monolayers.

\begin{table}[htbp]
\caption{\label{T1}The phonon PAM of six phonon modes at $K$ under vertical electric fields.}
\
\centering
\renewcommand\arraystretch{1.5}
\setlength{\tabcolsep}{0.5mm}
\begin{tabular}{c|cccccc}
\hline
\hline
&Mode 1 &Mode 2 &Mode 3 &Mode 4 &Mode 5 &Mode 6 \cr
\hline
  $E_{z}\!=\!+0.10$V/\AA & $-1$ & $+1$ & $0$ & $+1$ & $-1$ & $0$\cr
  $E_{z}\!=\!-0.10$V/\AA & $+1$ & $-1$ & $0$ & $-1$ & $+1$ & $0$\cr
\hline
\hline
\end{tabular}

\end{table}

\subsection{Discussions}

We investigate the gate-tunable chiral phonons at $K$ above.
There is another high-symmetry $-K$ point, which is related to the $K$ point by time-reversal symmetry \cite{xiao2007,xiao2012}.
 The computed phonon frequencies and gap evolutions are the same at $-K$ with those at $K$.
 In contrast, for phonon modes with the same frequency,
 the computed $S_{ph}^z$ and PAM at $-K$ are opposite to those at $K$,
 demonstrating contrasting phonon chiralites.
  Both the frequency degeneracy and opposite chiralities are ensured by time-reversal symmetry.

 Given that PAM can contribute a selection rule  \cite{zhang2015,zhu2018,zhang2020},
 the chiral-phonon-assisted optical transition can also be modulated by applied electric field.
 When PAM becomes opposite under reversed electric field,
 the chirality of applied optical field is expected to be changed correspondingly to enable the optical transition.
 Moreover, domains with opposite phonon chiralities are likely to be formed by applying opposite gates on different regions of a group-IVA monolayer,
 which will give rise to topological protected one-dimensional thermal transport channels along the domain boundary \cite{qiao2011,qiao2014}.
 By patterning opposite gates, a grid of thermal transport channels can be further fabricated
 and used to design nanoscale phononic and valley devices.
   In addition, two-dimensional ferroelectric materials are also worth study \cite{di2015},
since they have stronger response in structure to electric field that may further boost the phononic gap.

\section{Conclusions}
\label{sect4_conclu}

In summary, we performed first-principles calculations to investigate the effect of vertical electric fields
on chiral phonons in low-buckled group-IVA monolayers.
Since the vertical electric field breaks spatial inversion symmetry,
 phononic gaps are opened at high-symmetry $\pm K$ points of the momentum space.
 The magnitudes of phononic gaps increase linearly with the electric field strength,
 and group-IVA monolayers considered here exhibit different increasing rates.
 At $\pm K$,
 definite chiralities, characterized by considerable circular polarizations and discrete PAM,
 are generated for the gapped phonon modes.
The chiralities of phonons become opposite by turning around the direction of vertical electric fields.
The gate-tunable chiral phonons are expected to be involved in modulations of optical transitions and thermal transports,
offering opportunities for investigating couplings of chiral phonons with other quasiparticles and exploiting related devices.

\begin{acknowledgments}

We acknowledge supports from the National Natural Science Foundation of China (Nos. 11904173, 11975125, 12004186).
X.L. is also supported by the Jiangsu Specially-Appointed Professor Program.

\end{acknowledgments}

\bigskip
\
\appendix
\section{Phonon dispersions of germanene and stanene}
\label{appendix}

\begin{figure}[htbp]
\centerline{\includegraphics[width=8.5cm]{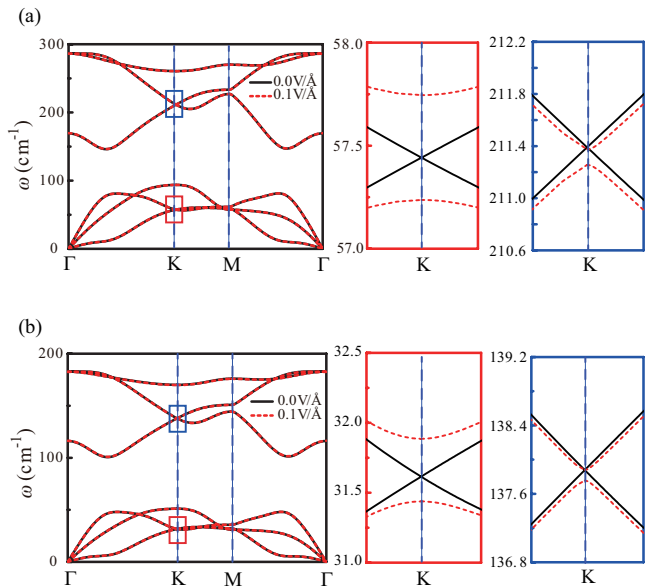}}
\caption{\label{GeSn}
 Phonon dispersions and magnified ones around $K$ before and after applying electric fields.
(a) The germanene monolayer.
(b) The stanene monolayer.
The detailed descriptions can refer to the caption of Fig. \ref{P2}.
 }
\end{figure}

Fig. \ref{GeSn} (a) and (b) show phonon dispersions of germanene and stanene, respectively.
Left panels of (a) and (b) correspond to phonon band structure along the path of $\Gamma-K-M-\Gamma$,
while right panels present the dispersion of acoustic and optical branches around $K$.
The degenerate modes at $K$ open gaps under vertical electric field, similar to the case of silicene.
 The gapped modes possess definite chirality with nonzero phonon circular polarizations, as shown in Fig. \ref{P5} of main text.

\bibliographystyle{unsrt}
\bibliography{chiral}

\end{document}